\documentclass[final]{siamltex} 
\usepackage{psfig,labelfig} 
\input amssymb.sty 
\newcommand{\eps}{\varepsilon} 
 
\newcommand{\iint}{\int \!\! \int} 
 
\newcommand{\dfrac}{\displaystyle\frac} 
\newcommand{\pdr}[2]{\dfrac{\partial{#1}}{\partial{#2}}} 
\newcommand{\bw}{{\bf w}} 
\newcommand{\br}{{\bf r}} 
\newcommand{\bS}{{\bf S}} 
\newcommand{\commentout}[1]{}   
\newcommand{\vx}{{\bf x}} 
\newcommand{\vy}{{\bf y}} 
 
\newcommand{\bp}{{\bf p}} 
\newcommand{\bk}{{\bf k}} 
\newcommand{\bq}{{\bf q}} 
 
\newcommand{\vX}{{\bf X}} 
 
\newcommand{\bP}{{\bf P}} 
\newcommand{\vB}{{\bf B}} 
\newcommand{\be}{\begin{equation}} 
\newcommand{\bQ}{{\bf Q}} 
\newcommand{\ee}{\end{equation}} 
\newcommand{\vY}{{\bf Y}} 
\newcommand{\vZ}{{\bf Z}} 
 
\newcommand{\PP}{{\mathord{I\kern -.33em P}}} 
\newcommand{\EE}{{\mathord{I\kern -.33em E}}}

\newcommand{\bZ}{{\bf Z}} 

\begin{document}

\title{Statistical stability in time 
  reversal} 
 
\author{George Papanicolaou \thanks{%
Department of Mathematics, Stanford University, Stanford CA, 94305; 
papanico@math.stanford.edu, 
} \and Leonid Ryzhik  
\thanks{Department of Mathematics, University of Chicago, Chicago IL, 60637; 
ryzhik@math.uchicago.edu, 
}\and Knut S\O lna \thanks{ 
Department of Mathematics, University of California, Irvine CA, 92697; 
ksolna@math.uci.edu. 
}} 
\maketitle  
\begin{abstract} 
When a signal is emitted from a source, 
recorded by an array of transducers, time reversed 
and re-emitted into the medium,  
it will refocus approximately on the  
source location. We analyze the refocusing resolution in a high 
frequency, remote sensing regime, and show 
that, because of multiple scattering, in an inhomogeneous or random medium 
it can improve beyond the diffraction limit.  
We also show that the back-propagated 
signal from a spatially localized narrow-band source 
is self-averaging, or statistically stable, and relate this 
to the self-averaging properties of functionals of the Wigner 
distribution in phase space. Time reversal 
from spatially distributed sources is 
self-averaging only for broad-band signals. 
The array of transducers operates in 
a remote-sensing regime so we analyze time reversal with the parabolic or 
paraxial wave equation. 
\end{abstract} 
 
 
 
\pagestyle{myheadings} 
\thispagestyle{plain} 
\markboth{G. PAPANICOLAOU, L. RYZHIK AND K. S\O LNA} 
{Parabolic approximation and time reversal}

\section{Introduction} 
 
\begin{figure}[tbh] 
\begin{center} 
\SetLabels 
\E  (-0.04*0.55)  $\frac{\lambda L}{a_e}$ \\ 
\E  (0.45*0.05)  $L$ \\ 
\E  (1.03*0.65)     $a$ \\ 
\endSetLabels 
\strut\AffixLabels{\psfig{file=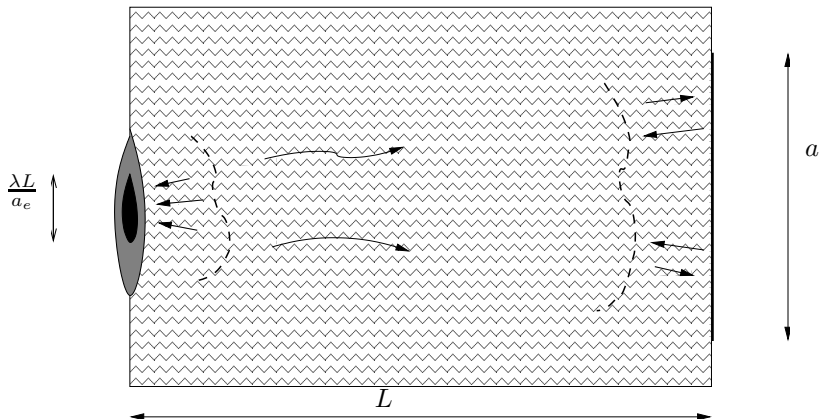,width=.7\hsize}} 
\end{center} 
\caption{ 
A pulse propagates toward a time 
reversal array of size $a$.  The propagation 
distance $L$ is large compared to $a$. 
The ambient medium has a randomly varying index of refraction 
with a typical correlation length that is 
small compared to $a$. 
The signal is time reversed at the array and sent back 
into the medium. The back propagated signal refocuses 
with spot size $\lambda L/a_e$, where $a_e$ is the
effective aperture of the array (Section \ref{mean}). 
} 
\label{figout} 
\end{figure}

In time reversal experiments a signal emitted by a localized source is 
recorded by an array and then re-emitted into the medium 
time-reversed, that is, the tail of the recorded signal is sent back 
first. In the absence of absorption the re-emitted signal propagates 
back toward the source and focuses approximately on it. This 
phenomenon has numerous applications in medicine, underwater acoustics 
and elsewhere and has been extensively studied in the literature, both 
from the experimental and theoretical points of view 
\cite{DJ90,DJ92,Fink-Nonlin,FCDPRTTW,Fink-Prada-01, 
HSK99,KHS97,KHSAFJ, fink2}. Recently time reversal has been also 
the subject of active mathematical research in the context of wave 
propagation and imaging in random media 
\cite{Bal-Ryzhik1,Bal-Ryzhik2,BPR,BPZ,BTPB-1,BTPB-2,TsP}. 
A schematic 
description of a time reversal experiment is presented in Figure \ref{figout}. 
 
For a point source in a homogeneous medium, the size of 
the refocused spot is approximately $\lambda L 
/a$, where $\lambda$ is the central wavelength of the emitted signal, 
$L$ is the distance between the source and the transducer array and 
$a$ is the size of the array. We assume here that the array is operating 
in the remote-sensing regime $a \ll L$.  
Multiple scattering in a randomly inhomogeneous medium 
creates {\bf multipathing}, which means that the transducer array can 
capture waves that were initially moving away from it but get 
scattered onto it by the inhomogeneities. As a result, the array 
captures a wider aperture of rays emanating from the original source 
and appears to be larger than its physical size. Therefore,  
somewhat contrary to intuition, 
the inhomogeneities of the medium do not destroy the refocusing 
but enhance its resolution. The refocused 
spot is now $\lambda L /a_e $, where $a_e > a$ is the {\bf effective} size of the array 
in the randomly scattering medium, and depends on $L$. The enhancement of refocusing 
resolution by multipathing is called {\bf super-resolution} \cite{BPZ}. 
The time reversed pulse is also 
{\bf self averaging} and refocusing near the source is therefore 
{\bf 
statistically stable}, which means that it does not depend on 
the particular realization of the random medium. There is some loss 
of energy in the refocused signal because of scattering away from the array 
but this can be overcome by amplification, up to a point. 
 
The purpose of this paper is to explore in detail the mathematical 
basis of pulse stabilization, beyond what was done in \cite{BPZ}. 
We want to explore in particular in what regime of parameters  
statistical stability is observed in time reversal. 
We show here that for high frequency waves in a remote sensing regime, 
spatially localized sources lead to statistically 
stable super-resolution in time reversal, even for narrow-band signals. 
We also show that 
when the source is spatially distributed, 
only for broad-band signals do we have  
statistical stability in time reversal. 
The regime where our analysis holds is a high frequency one, 
more appropriate to optical or infrared time reversal than to 
ultrasound, sonar or microwave radar. In this regime we 
can make precise what spatially localized or 
distributed means (see Section \ref{stab}). The numerical simulations 
in \cite{BPZ} and \cite{BTPB-1}, which are set 
in ultrasound or underwater sound regime, indicate that time reversal 
is not statistically stable for narrow-band signals even for 
localized sources. Only for broad-band signals is time reversal 
statistically stable in the regime of ultrasound experiments 
or sonar.

If the aperture of the transducer array is small $a/L \ll 1$, 
the Fresnel number ${L}/(k a^2)$ is of order one, 
and the random 
inhomogeneities are weak, which is often the case, we may analyze wave 
propagation in the paraxial or parabolic approximation \cite{Tappert}. 
The wave field is then given approximately by 
\begin{equation} 
\label{full-field} 
u(t,{\bf x},z)=  
\frac{1}{2\pi}\int e^{i\omega( z/c_0 - t )} \psi(z,{\bf x};\omega/c_0) d\omega 
\end{equation} 
where the complex amplitude $\psi$ satisfies the parabolic or 
Schr\"odinger equation 
\be 
\label{parabolic-1} 
2ik \psi_z + \Delta_{\bf x} \psi + k^2 (n^2 -1)\psi =0. 
\ee 
Here ${\bf x}=(x,y)$ are the coordinates transverse to the 
direction of propagation $z$, the wave number $k=\omega/c_0$ and 
$n({\bf x},z)=c_0/c({\bf x},z)$ is the random 
index of refraction relative to a reference speed $c_0$. 
The fluctuations of the refraction index 
\be 
\label{fluctuations} 
\sigma \mu(\frac{{\bf x}}{l},\frac{z}{l})=n^2({\bf x},z)-1 
\ee 
are assumed to be a stationary random field with mean zero, 
variance $\sigma^2$, correlation length $l$  and 
normalized covariance with dimensionless arguments 
\be 
\label{covariance} 
R({{\bf x}},{z})= 
E\{\mu({{\bf x}}+ {{\bf x}}',{z}+{z}') 
\mu({{\bf x}}',{z}') \} . 
\ee 
 
A convenient tool for the analysis of wave propagation in a random medium is 
the {Wigner distribution} \cite{GMMP,RPK-WM} defined by 
\be 
\label{wigner} 
W(z,\vx,\bp)=\frac{1}{(2\pi)^d}\int_{{\mathbb R}^d} 
e^{i\bp\cdot\vy}\psi(\vx-\frac{\vy}{2},z) 
\overline{\psi(\vx+\frac{\vy}{2},z)}d\vy 
\ee 
where $d=1$ or $2$ is the transverse dimension 
and the bar denotes complex conjugate.  The 
Wigner distribution may be interpreted as phase space wave energy 
and is particularly well suited for high frequency asymptotics and 
random media \cite{RPK-WM}.  The quantity of principal interest in 
time reversal, the time-reversed and back-propagated wave field, can 
be also expressed in terms of the Wigner distribution (see Section \ref{stab}). 
The self-averaging properties of the back-propagated field are related 
to the self-averaging properties of functionals of the Wigner distribution 
in the form of integrals of $ W$ over the wave numbers $\bp$. 
 
In the next Section we introduce a precise scaling that corresponds to 
(a) high frequency, (b) long propagation distance, (c) narrow beam 
propagation, and (d) weak random fluctuations. In the asymptotic limit 
where the small parameters go to zero the Wigner distribution 
satisfies a stochastic partial differential equation (SPDE), a 
Liouville-Ito equation, that has the form 
\be 
\label{ito-liouville} 
dW(z,\vx,\bp;k)=\left(-\frac{\bp}{k}\cdot\nabla_{\vx}W +\frac{k^2 D}{2} \Delta_{\bp} W 
\right) dz -\frac{k}{2} \nabla_{\bp}W \cdot d {\bf B}(\vx,z) 
\ee 
where ${\bf B}(\vx,z)$ is a vector-valued Brownian field with covariance 
\be 
\label{bf-covariance} 
E\{B_i(\vx_1,z_1)B_j(\vx_2,z_2)\}=-\left( 
\frac{\partial^2R_0((\vx_1-\vx_2))}{\partial x_i\partial x_j}\right) 
z_1 \wedge z_2, 
\ee 
where $z_1 \wedge z_2 = \min\{z_1 , z_2 \}$,  and in the isotropic case
\begin{equation} 
\label{D} 
D=-\frac{R_0^{''}(0)}{4} ,~~~R_0(\vx)=\int_{-\infty}^\infty R(\vx,s)ds. 
\end{equation} 
In Section \ref{sec:decorrelation} we analyze this SPDE in the 
asymptotic limit of small correlation length for ${\bf B}(\vx,z)$  
in the transverse variables $\vx$, 
and show that $W(z,\vx,\bp;k)$'s with different wave vectors $\bp$ are 
uncorrelated.  From this decorrelation property we deduce  
that for localized sources the time-reversed, 
back-propagated field is self-averaging, even for narrow-band signals. 
For distributed sources it is self-averaging only for broad-band 
signals. 
We show in detail in Section \ref{sec:TR} 
how the asymptotic theory is used in 
time reversal.  In Appendix \ref{add2} we introduce other scalings 
which lead to the same averaged SPDE but we do not analyze them 
in detail.  
 
Throughout the paper we define the Fourier transform by 
\[ 
\hat f(\bk)=\int d\vx e^{-i\bk\cdot\vx}f(\vx) 
\] 
so that 
\[ 
f(\vx)=\int \frac{d\bk}{(2\pi)^d} e^{i\bk\cdot\vx}\hat f(\bk). 
\] 
 
G. Papanicolaou was supported in part by grants AFOSR 
F49620-01-1-0465, NSF DMS-9971972 and ONR N00014-02-1-0088, L. Ryzhik 
by NSF grant DMS-9971742, an Alfred P. Sloan Fellowship and ONR grant 
N00014-02-1-0089.  K. Solna by NSF grant DMS-0093992 and ONR grant 
N00014-02-1-0090.

\section{Scaling and asymptotics} 
 
\subsection{The rescaled problem}\label{sec:scaling} 
 
To carry out the asymptotic analysis we begin by 
rewriting the Schr\"odinger equation (\ref{parabolic-1}) in  
dimensionless form. Let $L_z$ and $L_\vx$ be  characteristic 
length scales in the propagation direction, as, for example,
the distance $L$ between the source and 
the transducer array for $L_z$  
and the array size $a$ for $L_x$. 
We introduce a dimensionless wave number $k'=k/k_0$ 
with $k_0=\omega_0/c_0$ and $\omega_0$  a central 
frequency. We rescale $\vx$ and $z$ by 
$\vx=L_\vx\vx'$, $z=L_zz'$ and rewrite (\ref{parabolic-1}) in the new 
coordinates dropping primes: 
\begin{equation} 
\label{parabolic-1-rescaled} 
{2i}k\pdr{\psi}z + \frac{L_z}{k_0L_\vx^2}\Delta \psi +  
k^2 k_0L_z\sigma\mu\left(\frac{\vx L_\vx}{l},\frac{zL_z}{l}\right)\psi =0. 
\end{equation} 
 
The physical parameters that characterize the propagation 
problem are: (a) 
the central wave number $k_0$, (b) 
the strength of the fluctuations $\sigma$, and 
(c) the correlation length $l$. 
We introduce now three dimensionless variables 
\be 
\label{dimensionless-parameters} 
\delta=\frac{l}{L_\vx},~~ \eps=\frac{l}{L_z},~~\gamma=\frac{1}{k_0 l} 
\ee 
which are the reciprocals of 
the {\bf transverse scale} relative to correlation length, 
the reciprocal of the {\bf propagation distance} relative to correlation length, 
and the central {\bf wave length} 
relative to the correlation length. 
We will assume that the dimensionless parameters $\gamma$, $\sigma$, 
 $\eps$ and 
$\delta$ are small 
\be 
\label{order} 
\gamma ~\ll~ 1; ~~~ \sigma ~\ll~ 1; ~~~ \delta ~\ll~ 1; ~~~\eps ~\ll ~1 ~. 
\ee 
This is a regime of parameters where 
super-resolution phenomena can be observed.  
 
To make the scaling more precise we introduce 
the Fresnel number  
\be 
\label{fresnel} 
\theta =\frac{L_z}{k_0 L_\vx^2}=\gamma\frac{\delta^2}{\eps}    . 
\ee 
We can then rewrite 
the Schr\"odinger equation (\ref{parabolic-1-rescaled}) 
in the form 
\begin{eqnarray} 
  \label{eq:parabolic-rescaled} 
  &&2ik\theta\psi_z+\theta^2\Delta_\vx\psi+ 
\frac{k^2 \delta}{\eps^{1/2}}\mu(\frac{\vx}{\delta},\frac{z}{\eps})\psi=0. 
\end{eqnarray} 
provided that we relate $\eps$ to $\sigma$ and $\delta$ by 
\be 
\label{parameter-eps} 
\eps=\sigma^{2/3} \delta^{2/3} . 
\ee 
One way that the asymptotic regime  
(\ref{order}) can be realized is with the ordering 
\be 
\label{order-2} 
\theta ~\ll ~\eps ~\ll ~ \delta ~\ll 1 ~, 
\ee 
and $\gamma ~\ll~ \sigma^{4/3} \delta^{-2/3}$,  
corresponding to the high-frequency limit. 
We see from the scaled Schr\"odinger equation (\ref{eq:parabolic-rescaled}) 
that this regime can be given the following interpretation. We have first  
a {\bf high frequency} limit $\theta\to 0$, then a {\bf white noise} 
limit $\eps \to 0$, and then a {\bf broad beam} limit $\delta\to 0$. 
We will analyze in detail and interpret these limits in the 
following Sections. Another scaling  
in which (\ref{order}) is realized 
is $\eps ~\ll ~\theta ~\ll \delta ~\ll 1$. This is a regime in which 
the white noise limit is carried out first, then the high frequency 
limit and then the broad beam limit. We do not analyze this case 
here. 
Additional comments on scaling are provided in Appendix \ref{add2}.

It is instructive to express the constraints (\ref{parameter-eps}) and 
(\ref{order-2}) in terms of the dimensional parameters of the 
problem.  First, both the size of the transverse scale $L_\vx$ 
and the propagation distance $L_z$ should be much larger than the 
correlation length $l$ of the medium. Moreover, (\ref{parameter-eps}) 
implies that the longitudinal and transverse scales should be related 
by 
\[ 
\frac{L_z}{L_\vx}=\left(\frac{\delta}{\sigma^2}\right)^{1/3}\gg 1 
\] 
so that we are indeed in the beam approximation. The first inequality 
in (\ref{order-2}) implies that 
\[ 
\frac{L_z}{L_\vx}\ll \sqrt{k_0l} = \frac{1}{\sqrt{\gamma}}, 
\] 
and with the above choice of $L_z$ this implies that 
\[ 
\frac{\gamma^{3/2}}{\sigma^2}\ll \frac{L_\vx}{l}\ll \frac{1}{\sigma^2}. 
\]

\subsection{The high frequency limit} 
 
A convenient tool for the study of the high 
frequency limit, especially in random media, is the Wigner 
distribution.  It is often used in the context of energy propagation 
\cite{GMMP,RPK-WM} but it is also useful in analyzing 
time reversal phenomena \cite{Bal-Ryzhik1,Bal-Ryzhik2,BPZ}.  
Let $\phi_\theta(\vx)$ be a family of functions 
oscillating on a small scale $\theta$. The Wigner distribution is a function 
of the physical space coordinate $\vx$ and wave vector $\bp$ defined as 
\begin{equation}\label{eq:wig-def} 
W_\theta(\vx,\bp)=\int\limits_{{\mathbb 
R}^d}\frac{d\vy}{(2\pi)^d}e^{i\bp\cdot\vy} 
\phi_\theta(\vx-\frac{\theta\vy}{2}) 
\overline{\phi_\theta(\vx+\frac{\theta\vy}{2})}. 
\end{equation} 
The family $W_\theta$ is bounded in the space of Schwartz 
distributions ${\cal S}'({\mathbb R}^d\times{\mathbb R}^d)$ if the 
functions $\phi_\theta$ are uniformly bounded in $L^2({\mathbb 
R}^d)$. Therefore there exists a subsequence $\theta_k\to 0$ such that 
$W_{\theta_k}$ converges weakly as $k\to\infty$ to a limit measure 
$W(\vx,\bp)$. This limit $W(\vx,\bp)$ is non-negative and is customarily 
interpreted as the limit phase space energy density because 
\be 
\label{inten} 
|\phi_{\theta_k}(\vx)|^2\to \int\limits_{{\mathbb R}^d} W(\vx,\bp)d\bp 
~~~\hbox{as $\theta\to 0$} 
\ee 
in the weak sense. This allows one to think of $W(\vx,\bp)$ as a local 
energy density.  

Let $W_\theta(z,\vx,\bp)$ be the Wigner distribution of the solution 
$\psi$ of the Schr\"odinger equation (\ref{eq:parabolic-rescaled}), in 
the transversal space-variable $\vx$.  A straightforward calculation 
shows that $W_\theta(z,\vx,\bp)$ satisfies in a weak sense the linear 
evolution equation 
\begin{eqnarray} 
\label{W-transport} 
&&\pdr{W_\theta}{z}+\frac{\bp}{k}\cdot  
\nabla_\vx W_\theta \\ 
&&=\frac{ik\delta}{2\sqrt{\eps}} 
\int e^{i\bq\cdot\vx/\delta} 
\hat\mu\left(q,\frac{z}{\eps}\right) 
\frac{W_\theta\left(\bp-\frac{\theta\bq}{2\delta}\right)- 
W_\theta\left(\bp+\frac{\theta\bq}{2\delta}\right)}{\theta}  
\frac{d\bq}{(2\pi)^d}.\nonumber 
\end{eqnarray} 
In the limit $\theta \to 0$ the solution converges 
weakly in ${\cal S}'$, for each realization, 
to the (weak) solution of the random Liouville equation 
\be 
\label{random-liouville} 
\pdr{W}{z}+\frac{\bp}{k}\cdot 
\nabla_\vx W + 
\frac{k}{2\sqrt{\eps}}\nabla_\vx \mu\left(\frac{\vx}{\delta} 
,\frac{z}{\eps}\right)\cdot\nabla_\bp W =0 . 
\ee 
The initial condition at $z=0$ is $W(0,\vx,\bp)=W_I(\vx,\bp)$, the limit 
Wigner distribution of the initial wave function.   
 
\subsection{The white noise limit}\label{sec:one-point} 
 
In this Section we take the white noise limit $\eps\to 0$ in the 
random Liouville equation (\ref{random-liouville}) whose solution we 
now denote by $W_\eps$.  We can do this using the asymptotic theory of 
stochastic differential equations and flows 
\cite{Kohler-Papanico,BP78,KP79,K84} as follows.  
Using the method of characteristics, 
the solution of the 
Liouville equation (\ref{random-liouville}) may be written  
in the form 
\[ 
W_\eps(t,\vx,\bp)=W_I({\vX}_\eps(t;\vx,\bp),{\bP}_\eps(t;\vx,\bp)), 
\] 
where the processes $\vX_\eps(t;\vx,\bp)$ and $\bP_\eps(t;\vx,\bp)$ are 
solutions of the characteristic equations 
\[ 
\frac{d\vX_\eps}{dz}=-\frac{1}{k}\bP_\eps;~~~ 
\frac{d\bP_\eps}{dz}=-\frac{k}{2\sqrt{\eps}} 
\nabla_\vx \mu\left(\frac{\vX_\eps}{\delta},\frac{z}{\eps}\right) 
\] 
with the initial conditions $\vX_\eps(0)=\vx$ and $\bP_\eps(0)=\bp$. The 
asymptotic theory of random differential equations with rapidly 
oscillating coefficients implies that, under suitable conditions 
on $\mu$, in the 
limit $\eps\to 0$ the processes $\vX_\eps$, $\bP_\eps$ converge weakly 
(in the probabilistic sense), and uniformly on compact sets in 
$\vx,\bp$ to the limit processes $\vX(t)$, $\bP(t)$ that satisfy a 
system of stochastic differential equations 
\[ 
d\bP=-\frac{k}{2}d{\bf B}(z),~~~ 
d\vX=-\frac{1}{k}\bP dz,~~~\vX(0)=\vx,~~\bP(0)=\bp. 
\] 
The random process ${\bf B}(z)$ is a Brownian motion with the covariance 
function 
\begin{eqnarray}\label{one-cov} 
E\left\{B_i(z_1)B_j(z_2)\right\}&& = 
 -\frac{\partial^2R_0 (0)}{\partial x_i\partial x_j}ds 
z_1\wedge z_2\\ 
&&=\delta_{ij} 
\left(-R_0^{''}(0)\right) 
z_1\wedge z_2 , \nonumber 
\end{eqnarray} 
in the isotropic case, where 
\begin{equation} 
\label{rzero} 
R_0(\vx)=\int_{-\infty}^\infty R(\vx,s)ds 
\end{equation} 
is a function of $|\vx|$. 
This implies that the average Wigner distribution 
$W_\eps^{(1)}(z,\vx,\bp)=E\left\{W_\eps(z,\vx,\bp) \right\}$  
converges as $\eps\to 0$ uniformly on compact sets to the solution of the 
advection-diffusion equation in phase space 
\begin{equation}\label{eq:difflimit} 
\pdr{W^{(1)}}{z}+\frac{\bp}{k}\cdot \nabla_\vx W^{(1)} 
=\frac{k^2D}{2}\Delta_\bp W^{(1)} 
\end{equation} 
with the initial data $W^{(1)}(0,\vx,\bp)=W_I(\vx,\bp)$. 
Here the diffusion coefficient $D$ is given by 
\begin{equation}\label{diff-coeff} 
D=-\frac{R_0^{''}(0)}{4}.  
\end{equation} 
 
The one-point moments 
$E\left\{[W_\eps(z,\vx,\bp)]^N\right\}$ converge as $\eps\to 0$ to the 
functions $W^{(N)}(z,\vx,\bp)$ that satisfy the same equation 
(\ref{eq:difflimit}) but with the initial data 
$W^{(N)}(0,\vx,\bp)=[W_0(\vx,\bp)]^N$. This is similar to the spot 
dancing phenomenon \cite{Dawson-Papanico}, where all one-point moments are 
governed by the same Brownian motion.  In particular we have 
that  
\[ 
W^{(2)}(z,\vx,\bp)\ne \left[W^{(1)}(z,\vx,\bp)\right]^2 
\] 
so that the process $W_\eps$ does not converge to a deterministic 
one, in the strong sense pointwise.  
 
\subsection{Multi-point moment equations} 
  
As in the previous Section we may also 
study the white noise limit $\eps\to 0$ of the higher 
moments of $W_\eps(z,\vx,\bp)$ at different points 
\[ 
W_\eps^{(N)}(z,\vx^1,\dots,\vx^N,\bp^1,\dots,\bp^N)= 
E\left\{[W_\eps(z,\vx^1,\bp^1)]^{r_1}\cdot\dots\cdot 
[W_\eps(z,\vx^N,\bp^N)]^{r_N} \right\}. 
\] 
Here the points $(\vx^m,\bp^m)$ are all distinct, $(\vx^n,\bp^n)\ne 
(\vx^m,\bp^m)$. We may account for moments that have different powers 
of $W_\eps$ at different points by taking different powers $r_j$ 
of $W_\eps(\vx^j,\bp^j)$.   
 
We now consider the joint process 
$(\vX_\eps(z;\vx^m,\bp^m), \bP_\eps(z;\vx^m,\bp^m))$, $m=1,\dots ,N$. 
As $\eps\to 0$ it converges to the solution of the system of stochastic 
differential equations 
\begin{equation}\label{Npoint-sde} 
d\bP_i^m=-\frac{k}{2}\sum_{n=1}^N\sum_{j=1}^d 
\sigma_{ij}\left(\frac{\vX^m-\vX^n}{\delta}\right) 
d{B}_j^n(z),~~~ 
d\vX^m=-\frac{1}{k}\bP^mdz, 
\end{equation} 
with the initial conditions 
\[ 
\vX^m(0)=\vx^m,~~\bP^m(0)=\bp^m. 
\] 
The d-dimensional Brownian motions ${\bf B}^m$, $m=1,\dots ,N$ have 
the standard covariance tensor 
\[ 
E\left\{B_i^m(z_1)B_j^n(z_2)\right\}=\delta_{mn}\delta_{ij}z_1\wedge z_2,~~ 
i,j=1,\dots,d,~m,n=1,\dots,N. 
\] 
The symmetric tensor $\sigma_{ij}(\vx)$ is determined from  
\begin{equation}\label{multi-cov} 
\sum_{k=1}^N \sigma_{ik}(\vx)\sigma_{jk}(\vx)=- 
\left(\frac{\partial^2R_0(\vx)}{\partial x_i\partial x_j}\right). 
\end{equation} 
We assume that equation (\ref{multi-cov})  
has a solution that is differentiable in $\vx$, which 
is compatible with the fact that the matrix on the 
right is, by Bochner's theorem, non-negative definite.   
 
The moments $W_\eps^{(N)}$ converge as $\eps\to 0$ to the solution of 
the advection-diffusion equation 
\begin{eqnarray}\label{eq:difflimit-N} 
&&\pdr{W^{(N)}}{t}+\sum_{m=1}^N\frac{\bp^m}{k}\cdot\nabla_{\vx^m} W^{(N)} 
=\frac{k^2D}{2}\sum_{m=1}^N\Delta_{\bp^m} W^{(N)}\\ 
&&~~~~~~~~~- 
\frac{k^2}{4}\sum_{n,m=1\atop{n>m}}^N\sum_{i,j=1}^d 
\frac{\partial^2R_0((\vx^n-\vx^m)/\delta)}{\partial x_i\partial x_j} 
\frac{\partial^2W^{(N)}}{\partial p_i^n\partial p_j^m}\nonumber 
\end{eqnarray} 
with the initial data 
\[ 
W^{(N)}(0,\vx_1,,\dots,\vx^N,\bp^1,\dots,\bp^N)= 
[W_I(\vx^1,\bp^1)]^{r_1}\cdot\dots\cdot [W_I(\vx^N,\bp^N)]^{r_N}. 
\] 
 From (\ref{eq:difflimit-N}) we can  calculate  moments of 
functionals of $W_\eps$ of the form 
\[ 
W_{\eps,\phi}(z)=\int W_\eps(z,\vx,\bp)\phi(\vx,\bp)d\vx d\bp. 
\] 
For example, as $\eps\to 0$ we have that 
\[ 
E\left\{[W_{\eps,\phi}(z)]^2\right\}\to 
\int W^{(2)}(z,\vx_1,\bp_1,\vx_2,\bp_2)\phi(\vx_1,\bp_1)\phi(\vx_2,\bp_2) 
d\vx_1d\bp_1d\vx_2d\bp_2. 
\] 
 
A convenient way to deal with 
not only the limit of $N$-point moments but with the full 
limit process $W(z,\vx,\bp)$, at all points $\vx,\bp$ simultaneously,  
is provided by the theory of stochastic 
flows \cite{Kunita}. For this we need to show that $W_\eps(z,\vx,\bp)$ 
converges weakly (in the probabilistic sense) as $\eps\to 0$ to the 
process $W(z,\vx,\bp)$ that satisfies the stochastic partial 
differential equation 
\begin{equation}\label{eq:ito-liouville} 
dW_\delta=\left[-\frac{\bp}{k}\cdot\nabla_\vx W_\delta + 
 \frac{k^2D}{2}\Delta_\bp W_\delta 
\right]dz-\frac k2\nabla_\bp W_\delta \cdot d{\bf B}(\frac{\vx}{\delta},z). 
\end{equation} 
Here the Gaussian random field ${\bf B}(\vx,z)$ has the covariance 
\[ 
E\{B_i(\vx_1,z_1)B_j(\vx_2,z_2)\}=-\left( 
\frac{\partial^2R_0((\vx_1-\vx_2))}{\partial x_i\partial x_j}\right) 
z_1 \wedge z_2. 
\] 
We call equation (\ref{eq:ito-liouville}) the  
Liouville-Ito equation. It allows us to treat all equations of the form 
(\ref{eq:difflimit-N}) simultaneously and is a convenient tool 
for simulation and analysis. The dimensionless 
wave number $k$ can be scaled out of (\ref{eq:ito-liouville}) by 
writing $W(z,\vx,\bp;k)=W(z,\vx,\frac{\bp}{k}; 1)$  
so that we need only consider (\ref{eq:ito-liouville}) with $k=1$. 
We will use this scaling in Section \ref{stab}. 
 
Note that unlike the single Brownian motion (\ref{one-cov}) that governs the 
evolution of one-point moments, the Brownian field that enters the SPDE 
(\ref{eq:ito-liouville}) 
depends explicitly on the dimensionless correlation length $\delta$ in 
the transverse direction. Therefore the limit process also depends on 
$\delta$ and we denote it by $W_\delta$.

\subsection{Statistical stability in the broad beam limit} 
\label{sec:decorrelation} 
 
We will now consider the limit $\delta\to 0$ of the process 
$W_\delta(z,\vx,\bp)$ when the transverse dimension $d\ge 2$.  We are 
particularly interested in the behavior of functionals of 
$W_{\delta}$ as $\delta\to 0$. The analysis of  
one-point moments in Section \ref{sec:one-point} showed 
that they do not depend on $\delta$ and are governed by a standard Brownian 
motion. Therefore the process $W_\delta$ does not have a pointwise 
deterministic limit. However, we will show that functionals of 
$W_\delta$ become deterministic in the limit $\delta\to 0$.  We refer 
to this phenomenon as {\bf statistical stabilization} and give 
conditions for it to happen. Stabilization plays an important 
role in time reversal, imaging and other applications, as discussed 
in the Introduction. 
\begin{theorem} 
\label{lem1} 
Assume that $\phi(\bp)$ is a smooth test function of rapid decay, 
the transverse correlation function $R_0(\vx)$ has compact support,  
the initial Wigner distribution 
 $W_I(\vx,\bp)$ is uniformly bounded and Lipschitz continuous, and the 
transverse dimension $d\geq 2$. Define 
\be 
\label{func} 
I_{\delta,\phi}(z,\vx) = \int W_\delta(z,\vx,\bp) \phi(\bp) d\bp. 
\ee 
Then 
\be\label{variance=0} 
  \lim_{\delta \to 0}  
E\left\{ I_{\delta,\phi}^2(z,\vx) \right\}= 
   E^2\left\{ I_{\delta,\phi}(z,\vx) \right\} 
\ee 
where $E\left\{ I_{\delta,\phi}(z,\vx) \right\}$ is independent of 
$\delta$. 
\end{theorem} 
 
The assumption of compact support for $R_0(\vx)$ is not essential 
but simplifies the proof.   
We have already noted that the Wigner 
distribution $W_\delta$ itself does not stabilize.   
However, (\ref{variance=0}) implies that 
\be 
  \lim_{\delta \to 0} 
Var\left\{ I_{\delta,\phi} \right\} = 
  \lim_{\delta \to 0} 
E\left\{ I_{\delta,\phi}^2(z) \right\}   
   - E^2 \left\{ I_{\delta,\phi} \right\} =0. 
\ee 
Therefore, any smooth functional 
of the form (\ref{func}) stabilizes in the limit $\delta \to 0$, 
that is, 
\be 
 I_{\delta,\phi} \approx E\{I_{\delta,\phi}\}, 
\ee 
in mean square, and the expectation of $I_{\delta,\phi}$ does not depend on $\delta$. 
We prove Theorem \ref{lem1} in Appendix \ref{app-c}.   
 
In the applications of the asymptotic theory to time reversal we need 
not only functionals $I_{\delta,\phi}$ of the form (\ref{func}) but also 
of the form 
\be 
\label{func2} 
J_{\delta}(z,\vx) =\int W_{\delta} (z,\vx,\bp) d\bp . 
\ee 
We need to show that such functionals are well defined with 
probability one and to analyze their behavior as $\delta \to 0$. 
This is done in the following theorem.   
 
\begin{theorem} 
\label{lem2} 
Under the same hypotheses of Theorem \ref{lem1} and with 
a non-negative initial Wigner distribution $W_I \geq 0$, 
the functional $J_{\delta}$ is bounded, continuous and non-negative 
with probability one. In the limit $\delta\to 0$ we have 
\be\label{J-variance=0} 
  \lim_{\delta \to 0} 
E\left\{ J_{\delta}^2(z,\vx) \right\}= 
   E^2 \left\{ J_{\delta}(z,\vx) \right\} 
\ee 
where $E\left\{ J_{\delta}(z,\vx) \right\}$ does not depend on $\delta$. 
\end{theorem} 
 
The proof of this theorem is given in Appendix \ref{app-c}. 
 
What is important in both Theorems \ref{lem1} and \ref{lem2} is that 
we do integrate over the wave numbers $\bp$ because there is no 
pointwise stabilization. In time reversal applications, as in 
section \ref{stab}, we actually need Theorem \ref{lem2} when 
the integration is only over a line segment in $\bp$ space, 
and the dimension of the latter is $d\geq 2$. Its proof follows from the one 
of Theorem \ref{lem2}.

\section{Application to time reversal in a random medium}\label{sec:TR} 
 
We will now apply these results to the time reversal 
problem \cite{BPZ} described in the Introduction.   
A wave emitted from the plane $z=0$ propagates 
through the random medium and is recorded on the time reversal mirror at $L$. 
It is then {\em reversed} in time and re-emitted into the medium.  The 
back-propagated signal refocuses approximately at the source, as shown 
in Figure \ref{figout}. 
There are two striking features of this refocusing in random media. 
One is that  
it is statistically stable, that is, it does not 
depend on the particular realization.   
The other is super-resolution, that is, the refocused spot is tighter 
than in the deterministic case. We discuss these two issues in this 
section. 
 
\subsection{The time-reversed and back-propagated field} 
\label{stab} 
 
We assume that the wave source at $z=0$ 
is distributed on a scale $\sigma_s$  
around a point $\vx_0$, that is, 
\[ 
\psi_\theta(z=0,\vx;k) =   
{ e^{i\bp_0\cdot (\vx-\vx_0)/\theta} \psi_0(\frac{\vx-\vx_0}{\sigma_s};k)},  
\] 
where $\psi_0$ is a rapidly decaying and smooth function of $\vx$ and $k$. 
The width of the source $\sigma_s$ could be large or small compared 
to the  Fresnel number $\theta$, and this affects  
the statistical stability of the time-reversed, back-propagated 
field, as we explain in this Section. 
The Green's function, $G_\theta(z,\vx;\xi)$, 
solves the parabolic wave equation 
(\ref{eq:parabolic-rescaled}) with a point source at 
$(\vx,z)=(\xi,0)$. 
Using its symmetry properties and  
the fact that time reversal $t\to -t$ is equivalent to 
 $\omega\to -\omega$ or $k\to -k$,  the 
back-propagated, time-reversed field on the plane of the 
source has the form  
\begin{eqnarray} 
\label{eq:gpro} 
&&\psi_\theta^{B}(L,\vx_0,\xi;k)= \\ 
&& \iint  
G_\theta(L,\vx; \vx_0 + \theta\xi;k) 
\overline{ G_\theta(L,\vx_0 +  \eta ;\eta;k) } 
 e^{i \bp_0\cdot \eta /\theta }\psi_0(\frac{\eta}{\sigma_s};-k) \chi_A(\vx) d\vx d\eta.   
  \nonumber 
\end{eqnarray} 
The complex field amplitude $\psi_\theta^B$ is evaluated  
at $\vx_0+\theta\xi$, in the plane $z=0$. We scale the observation 
point off $\vx_0$ by $\theta$ because we 
expect that the spot size of the refocused signal will be comparable 
to the lateral spread of the initial wave function. 
We denote with $\chi_A$ the aperture function of the 
time reversal mirror. It could be its characteristic function, 
occupying the region $A$ in the plane $z=L$  
\begin{eqnarray*} 
  \chi_A(\vx)=\left\{\begin{matrix}{1, & \vx\in A\cr 0, & \vx \notin A\cr} 
\end{matrix}\right.  , 
\end{eqnarray*} 
or a  more general aperture function like a Gaussian. 
The time reversal mirror is located in the plane $z=L$. 
 
After changing variables, the back-propagated field is given by 
\begin{eqnarray*} 
 \psi_\theta^{B}(L,\vx_0,\xi;k)&=&  
  \theta^d \int  
G_\theta(L,\vx; \vx_0 + \theta\xi;k) 
\overline{ G_\theta(L,\vx;\vx_0 + \theta \eta;k) } 
e^{i\bp_0\cdot \eta}\psi_0(\frac{\theta\eta}{\sigma_s};-k) \chi_A(\vx) d\vx d\eta  \\ 
&=&  \theta^d \int  
G_\theta(L,\vx_0 + \theta\xi,\vx;k) 
\overline{ G_\theta(L,\vx_0 + \theta\eta,\vx;k) } 
 e^{i\bp_0\cdot \eta}\psi_0(\frac{\theta\eta}{\sigma_s};-k) \chi_A(\vx) d\vx d\eta . 
\end{eqnarray*} 
It is now convenient to introduce the Wigner distribution 
\begin{eqnarray} 
\label{Gwigner} 
\hbox{} ~ W_\theta(z,\vx_0,\bp;k)= \int \frac{\theta^d e^{i \bp \cdot \vy}} 
{(2\pi)^d}  
G_\theta(z,\vx_0  - \vy \theta/2,\vx;k) 
\overline{ G_\theta(z,\vx_0 + \vy \theta/2,\vx;k) } 
\chi_A(\vx) d\vx d\vy , 
\end{eqnarray}  
and express the back-propagated field as  
\begin{eqnarray} 
  \label{eq:back-prop} 
\psi_\theta^{B}(L,\vx_0,\xi;k) 
 &=&  \int e^{i\bp\cdot(\xi-\eta)} 
W_\theta(L,\vx_0+\frac{\theta(\xi+\eta)}{2},\bp;k) 
 e^{i\bp_0\cdot \eta}\psi_0(\frac{\theta\eta}{\sigma_s};-k) {d\bp d\eta}. 
\end{eqnarray} 
The Wigner distribution is scaled here differently from 
(\ref{eq:wig-def}) because of the way we have scaled the source 
function. 
 
In the high frequency limit $\theta \to 0$,  
 $W_{\theta}(z,\vx,\bp;k)$ tends to  $W(z,\vx,\bp;k)$, 
which solves the random Liouville 
equation (\ref{random-liouville}).  Then, in the 
white noise limit, it solves  
the Liouville-Ito equation (\ref{eq:ito-liouville}).  The mean of $W$ 
solves (\ref{eq:difflimit}), in the high-frequency and white noise limit,   
with initial data  
\be 
\label{IWF} 
W(0,\vx,\bp;k)=\frac{\chi_A(\vx)}{(2\pi)^d}. 
\ee 
Let  
\be 
\label{beta} 
\beta = \frac{\sigma_s}{\theta} 
\ee 
be the ratio of the width of the source to the Fresnel number 
and assume that it remains fixed as $\theta \to 0$. 
In this limit, the time-reversed and back-propagated 
field is given by 
\begin{eqnarray} 
\label{hfl} 
  \psi^{B}(L,\vx_0,\xi;k) 
& = & \int e^{i\bp\cdot(\xi-\eta)} 
 W(L,\vx_0,\frac{\bp}{k})  
e^{i\bp_0\cdot \eta}\psi_0({\eta}/{\beta};-k) {d\bp d\eta}\\ 
&=&\int e^{i\bp\cdot \xi} 
 W(L,\vx_0,\frac{\bp}{k})  
 \beta^d \hat\psi_0(\beta (\bp-\bp_0);-k) {d\bp} .  \nonumber 
\end{eqnarray} 
Here we have used the scaling $W(z,\vx,\bp;k)=W(z,\vx,\frac{\bp}{k}; 1)$ 
in (\ref{eq:ito-liouville}) and we have dropped the last 
argument $k=1$. 
 
\subsection{Statistical stability} 
\label{real-stab} 
 
From the form (\ref{hfl}) of the back-propagated and time-reversed 
field we see that when $\beta = O(1)$ (or small), which means that 
$\sigma_s$ is comparable to the Fresnel number $\theta$ (or smaller), 
we can apply the results of Section \ref{sec:decorrelation} and 
conclude that it is statistically stable or self-averaging in the 
broad beam limit $\delta \to 0$.  Theorems \ref{lem1} and \ref{lem2} 
are exactly what is needed for this.  The fact that the initial 
function (\ref{IWF}) may be discontinuous at the boundary of the set 
$A$ is not a problem. This is because, we may approximate the function $\chi_A$ 
from above and below by two smooth positive functions, to which we may 
apply Theorems \ref{lem1} and \ref{lem2}, and then use the maximum 
principle to deduce the decorrelation property when the initial data 
is $\chi_A$. 
\commentout{ 
 because we are in the white noise limit so we can 
replace the initial $W$ by a smooth one after advancing over a small 
interval $\Delta z$. Another way to avoid this difficulty is the 
formulate the expression (\ref{hfl}) for $\psi^B$ with the role of 
$\chi_A$ and $\psi_0$ reversed, as was done in \cite{BPZ}. 
}
 We have, 
therefore, 
\[ 
\psi^{B}(L,\vx_0,\xi;k) \approx \langle\psi^{B}(L,\vx_0,\xi;k)\rangle 
\] 
in the sense of convergence in probability or in mean square, 
in the broad beam limit $\delta \to 0$, 
for each  fixed frequency $\omega=kc_0$. Statistical stability of time reversal 
 does not 
depend on having a broad-band signal if the source is localized in space. 
This is true in the regime of parameters reflected by the scaling  
$\theta \ll \eps \ll \delta$ considered here,  
which is a high frequency regime encountered in 
optical or infrared applications like ladar. The numerical experiments in \cite{BPZ} 
and \cite{BTPB-1} are closer to the regime of ultrasound experiments 
\cite{Fink-Prada-01} and in underwater sound 
propagation, which is different from the high frequency regime analyzed here. 
 
For distributed sources the parameter $\beta$ is large and we cannot 
apply Theorems \ref{lem1} and \ref{lem2} to (\ref{hfl}). It is necessary 
for statistical stability in this case to have broad-band signals.  
For $\beta$ large  
the time reversed and back propagated signal in the time domain has 
the form 
\begin{eqnarray} 
\label{TR-time-domain} 
&&\psi^B(L,\vx_0,\xi,t) \\ 
&&=(2\pi)^d e^{i(\bp_0\cdot\xi-k_0 c_0 t)}\psi_0({\xi}/{\beta}) 
\int W(L,\vx_0,\frac{\bp_0}{k_0 + k}) e^{-ikc_0 t} \hat{g}(-c_0k) \frac{c_0 dk}{2\pi}  
\nonumber \\ 
&&=(2\pi)^d e^{i(\bp_0\cdot\xi-\omega_0 t)}\psi_0({\xi}/{\beta}) 
\int W(L,\vx_0,\frac{c_0\bp_0}{\omega_0 + \omega})  
e^{-i\omega t} \hat{g}(-\omega) \frac{ d\omega}{2\pi}  
\nonumber 
\end{eqnarray} 
with $\hat{g}(c_0 k)$ the Fourier transform of the initial pulse relative to the 
central frequency $\omega_0=c_0 k_0$. 
This means that we have replaced the actual wave number $k$ by $k_0 +k$, 
or $\omega$ by $\omega_0 +\omega$, 
with the new $\omega$, the baseband frequency, bounded by $\Omega$, the bandwidth, 
$|\omega|\leq \Omega < \omega_0$. The integration is over the 
bandwidth $[-\Omega,\Omega]$. 
This integral is well defined with probability one and is self-averaging 
in the broad beam limit $\delta \to 0$ by Theorem \ref{lem2} and the remark 
following it. We will compute its average in Section \ref{BB}.

\subsection{The effective aperture of the array} 
\label{mean} 
 
From the explicit expression for the Green's function 
of  (\ref{eq:difflimit}), with $k=1$,  
\begin{eqnarray*} 
&&U(z,\vx,\bp;\vx^0,\bp^0)=\int\frac{d\bw d\br}{(2\pi)^{2d}} 
\exp\left(i\bw\cdot(\vx-\vx^0)+i\br\cdot(\bp-\bp^0)-iz\bw\cdot\bp^0\right)\\ 
&&~~~~~~~~~~~~~~~~\times\exp\left(- 
\frac{Dz}{2}\left[r^2+z\br\cdot\bw+\frac{\bw^2z^2}{3}\right]\right), 
\end{eqnarray*} 
and with the time reversal mirror a distance $L$ from 
the source and $\vx_0 =0$, it follows from (\ref{hfl}) that 
\begin{eqnarray}\label{psib-lim} 
&&\langle\psi^{B}(z,\xi;k)\rangle =  \\ 
&&\int \frac{d\bp d\vy d\bw}{(2\pi)^{2d}} e^{i\bp\cdot\xi}  
\beta^d \psi_0(\beta(\bp-\bp_0);-k) \chi_A(\vy) \exp\left[-i\bw\cdot\vy- 
iz\bw\cdot\frac{\bp}{k}-\frac{Dz^3w^2}{6}\right] \nonumber 
\end{eqnarray} 
The high-frequency, white-noise limit of the {\em self-averaging} time-reversed and 
back-propagated field is therefore given by a convolution 
\begin{eqnarray} 
\label{w-conv} 
&&\langle\psi^B(L,\xi;k)\rangle=\psi_0^{\beta}(\cdot,-k) * {\cal W}(\cdot) (\xi)  
\end{eqnarray} 
with  
\begin{equation} 
\label{w-kernel} 
{\cal W}(\eta)={\cal W}(\eta;L,k)=\frac{k^d}{(2\pi L)^d}  
\hat \chi_A(\eta k/L)~ e^{-\eta^2 / (2 \sigma_M^2)} , 
\end{equation} 
the {\bf point spread function}, and 
\be 
\label{psi-beta} 
\psi_0^{\beta}(\eta,-k) =e^{i\bp_0\cdot\eta} 
\psi_0({\eta}/{\beta})\hat{g}(-kc_0) 
\ee 
with $\psi_0(\eta/\beta)$ the spatial source distribution function and  
$\hat{g}$ the Fourier transform of the pulse 
shape function $g(t)$. This notation is consistent with (\ref{TR-time-domain}), 
with the time factor $e^{-ik_0c_0 t}$ omitted, along with the horizontal 
phase $e^{ikz}$ which 
cancels in time reversal. 
We have also introduced the refocused {\bf spot size} with multipathing 
\be 
\label{variance-ae} 
\sigma_M^2 = \frac{3}{DL k^2} = \frac{L^2}{k^2 a_e^2} 
\ee 
and the {\bf effective aperture} $ a_e=a_e(L)$,  
\be 
\label{effective-aperture} 
a_e =\sqrt{\frac{DL^3}{3}}, 
\ee 
which we now interpret.

If the time reversal mirror is the whole plane $z=L$, then $\chi_A \equiv 1$ and 
\[ 
\left<\psi^B(L,\xi;k)\right>= 
\psi_0^{\beta}(\xi,-k) . 
\] 
In this case 
the back-propagated field is   
the source field reversed in time, 
both in the random and in the deterministic case. 
The point spread function ${\cal W}$  
determines the resolution of the refocused signal for a time 
reversal mirror of finite aperture. 
Multipathing in a random medium gives rise to the Gaussian 
factor (\ref{variance-ae})
whose variance is $\sigma_M^2$. 
We can give an interpretation of this variance, or spot size, as follows. 
For a square time reversal mirror of size $a$, the Fourier 
transform of $\chi_A$ is the sinc function 
so that 
\[ 
{\cal W}(\eta_1, \eta_2;L,k)= 
\left(\frac{1}{\pi L}\right)^2  \sin (\frac{\eta_1 k a}{2L})  
 \sin (\frac{\eta_2 k a}{2L})  
e^{-(\eta_1^2+\eta_2^2)/(2\sigma_M^2)} 
\] 
For a deterministic medium ($D=0$) the Rayleigh resolution is the distance $\eta_F$ to 
the first zero of the sine, the first Fresnel zone in either direction, 
\[ 
\eta_F = \frac{2\pi L}{k a} =\frac{\lambda L}{a} . 
\] 
In general, if $\chi_A$ is supported by a region of size $a$ 
we may define the Fresnel resolution, or the Fresnel {\bf spot size}, by 
\[ 
\sigma_F = \frac{L}{ka}. 
\] 
 
For {\bf weak multipathing} we have $\sigma_M \gg \sigma_F$ and 
\[ 
{\cal W}(\eta;L,k) \sim \left(\frac{k}{2\pi L}\right)^d 
\hat \chi_A(\eta k/L)~,  
\] 
which is the diffractive point spread function whose integral 
over $\eta \in R^d$ is one. 
If, however, we have {\bf strong multipathing}, $\sigma_M \ll \sigma_F$, 
then we may approximate $\hat \chi_A(\eta k/L)$ by $ \hat \chi_A(0) = a^{d}$ 
in (\ref{w-kernel}), and the point spread function becomes 
\[ 
{\cal W}(\eta;L,k) \sim \left(\frac{ka}{2\pi L}\right)^d 
e^{-|\eta|^2/(2\sigma_M^2)}. 
\] 
By writing the variance (spot size) $\sigma_M^2$  
in the form (\ref{variance-ae}) we can interpret $a_e$ as 
an effective aperture of the time reversal mirror. 
We can rewrite the point spread function in terms of a normalized Gaussian as 
\[ 
{\cal W}(\eta;L,k) \sim  \left(\frac{\sigma_M}{\sqrt{2\pi}\sigma_F}\right)^d  
\frac{e^{-|\eta|^2/(2\sigma_M^2)}}{(2\pi \sigma_M^2)^{d/2}} 
\] 
with the factor in front of the normalized Gaussian also equal to 
\[ 
\left(\frac{a}{\sqrt{2\pi}a_e}\right)^d. 
\] 
This means that when there is strong multipathing the integral 
of the point spread function over $R^d$ is not equal to one but to this 
ratio, which can be much smaller than one if $a_e \gg a$. Multipathing 
produces a tighter point spread function but there is also  
loss of energy, as of course we should expect. 
 
A more direct interpretation for the effective 
aperture can be given if the time reversal mirror has 
a Gaussian aperture function 
\begin{eqnarray*} 
\chi_A(\eta)=e^{-|\eta|^2/(2 a^2)}.  
\end{eqnarray*} 
 The point spread function $\cal W$ has now the form 
\[ 
{\cal W}(\eta;L,k) =  \left( \frac{ka}{\sqrt{2\pi} L}\right)^d  
   e^{- |\eta|^2 /(2 \sigma_{g}^2)},  
\] 
with  
\[ 
\sigma_{g}= \frac{L}{ka_{g}} 
\] 
and the effective aperture $a_{g}$ given by 
\[ 
a_{g} = \sqrt{a^2 + \frac{DL^3}{3}} =\sqrt{a^2 + a_e^2}. 
\] 
Clearly, $a_g \approx a_e$ when there is strong multipathing and 
$a_e \gg a$. 
Written with a normalized Gaussian the point spread function 
for a Gaussian aperture has the form 
\[ 
{\cal W}(\eta) =  \left( \frac{a}{a_g}\right)^d  
   \frac{e^{- |\eta|^2 /(2 \sigma_{g}^2)} }{ 
(2\pi \sigma_g^2)^{d/2}} . 
\]

%

\subsection{Broad-band time reversal for distributed sources} 
\label{BB} 
 
\begin{figure}[tbh] 
\begin{center} 
\SetLabels 
\E  (0.4*0.05)  $L$ \\ 
\E  (0.0*0.54)   $\sigma_s$ \\ 
\E  (1.02*0.67)   $a$ \\ 
\E  (0.9*0.67)   $\frac{|\bp_0| L}{k_0}$ \\ 
\endSetLabels 
\strut\AffixLabels{\psfig{file=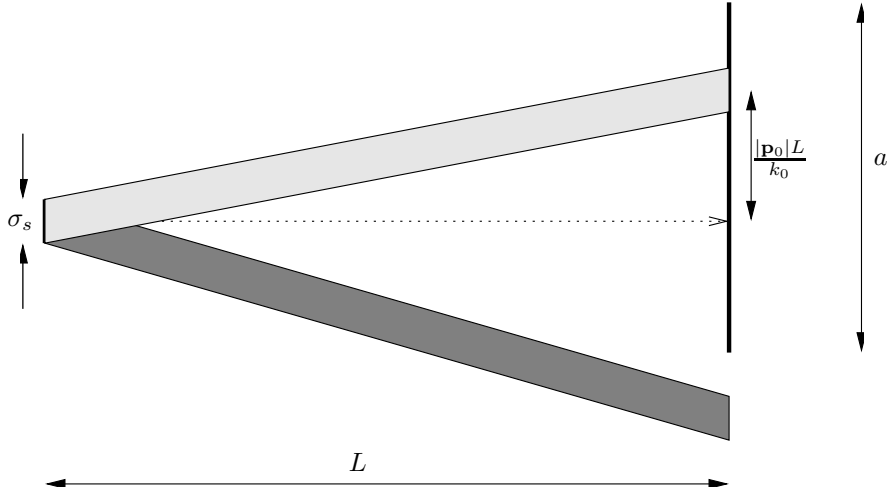,width=.8\hsize}} 
\end{center} 
\caption{ 
A directed field propagates from a distributed source 
of size $\sigma_s$ toward the time reversal mirror 
of size $a$. The time-reversed, back-propagated field 
depends on the location of the mirror relative 
to the direction of the propagating beam. 
} 
\label{figp0} 
\end{figure}

For a distributed source, its support $\sigma_s$ is large 
compared to the Fresnel number $\theta$ so the 
ratio $\beta=\sigma_s/\theta$ is large. In this case   
we can compute the average of (\ref{TR-time-domain}) the same way 
as we did in (\ref{psib-lim}) and we find that 
\begin{eqnarray} 
\label{Psib-lim} 
&&\langle \Psi^B(L,\vx_0,\xi,t) \rangle \\ 
&&=(2\pi)^d e^{i(\bp_0\cdot\xi-k_0 c_0 t)}\psi_0({\xi}/{\beta}) 
\int \langle W(L,\vx_0,\frac{\bp_0}{k_0 + k}) \rangle 
e^{-ikc_0 t} \hat{g}(-c_0 k) \frac{c_0 dk}{2\pi} \nonumber \\ 
&&=e^{i(\bp_0\cdot\xi-k_0 c_0 t)}\psi_0({\xi}/{\beta}) 
\int \frac{d\vy d\bw c_0 dk}{(2\pi)^{d+1}} \chi_A(\vy) 
e^{i(\frac{L\bw\bp_0}{k_0 +k} -\bw\cdot\vy - kc_0 t)} e^{-\frac{DL^3 w^2}{6}}  
\hat{g}(-c_0 k) . 
\nonumber 
\end{eqnarray} 
The $\vy$ integral on the right gives the Fourier transform of 
the aperture function $\chi_A(\vy)$ so with $\omega_0=c_0 k_0$ 
and a change of variable from $k$ to $\omega=c_0 k$ we have 
\begin{eqnarray} 
\label{Psib-lim-final} 
&&\langle \Psi^B(L,\vx_0,\xi,t) \rangle  \\ 
&&= e^{i(\bp_0\cdot\xi-\omega_0 t)}\psi_0({\xi}/{\beta}) 
\int \frac{ d\omega}{2\pi} e^{-i\omega t} \hat{g}(-\omega) 
 ~\chi_A * \left(\frac{e^{-x^2/(2a_e^2)}}{(2\pi a_e^2)^{d/2}}\right) 
(\frac{Lc_0\bp_0}{\omega_0 + \omega})  \nonumber 
\end{eqnarray} 
Here the star denotes convolution with respect to the 
spatial variables $\vx$, and $a_e$ is the effective 
aperture defined by (\ref{effective-aperture}).

When multipathing is weak we can 
ignore the Gaussian factor in the convolution and we have 
\begin{eqnarray} 
\label{Psib-lim-det} 
&&\langle \Psi^B(L,\vx_0,\xi,t) \rangle  \\ 
&&= e^{i(\bp_0\cdot\xi-\omega_0 t)}\psi_0({\xi}/{\beta}) 
\int \frac{d\omega}{2\pi} e^{-i\omega t} \hat{g}(-\omega) 
 ~\chi_A  
\left(\frac{Lc_0\bp_0}{\omega_0 + \omega} \right)  \nonumber 
\end{eqnarray} 
In the opposite case, when there is strong multipathing and 
the effective aperture is much larger than the physical one, 
$a_e \gg a$,  
we have 
\begin{eqnarray} 
\label{Psib-lim-rand} 
&&\langle \Psi^B(L,\vx_0,\xi,t) \rangle  \\ 
&&= e^{i(\bp_0\cdot\xi-\omega_0 t)}\psi_0({\xi}/{\beta}) 
\left(\frac{a}{\sqrt{2\pi}a_e}\right)^d  
\int \frac{d\omega}{2\pi} e^{-i\omega t} \hat{g}(-\omega) 
{e^{-\frac{1}{2}(\frac{Lc_0\bp_0}{a_e(\omega_0 + \omega)})^2}} 
  \nonumber 
\end{eqnarray}

To interpret these results we note first that a distributed 
source function of the form (\ref{psi-beta}) can be considered 
as a phased array emitting an inhomogeneous plane wave, a beam, in 
the direction $(k,\bp_0)$, within the paraxial or 
parabolic approximation. The ratio $|\bp_0|/k$ is the tangent 
of the angle the direction vector makes with the  
$z$ axis, and $L|\bp_0|/k$ is the transverse distance of the beam 
center to the center of the phased array (see Figure \ref{figp0}).  
If for each $\omega$ the beam 
displacement vector $Lc_0 \bp_0/(\omega_0 +\omega)$ is inside 
the set $A$ occupied by the time reversal array, then 
we recover at the source 
the full pulse 
in (\ref{Psib-lim-det}), 
time-reversed, 
\[ 
\langle \Psi^B(L,\vx_0,\xi,t) \rangle = 
e^{i(\bp_0\cdot\xi-\omega_0 t)}\psi_0({\xi}/{\beta}) g(-t) . 
\] 
If, however, for some frequencies the transverse displacement 
vector is outside the time reversal array, these frequencies will be nulled 
in the integration and a distorted time pulse will be 
received at the source. Depending on the position of 
the time reversal mirror relative to the beam, high or low 
frequencies may be nulled. 
 
In a strongly multipathing medium the situation is quite 
different because the expression (\ref{Psib-lim-rand}), 
or more generally (\ref{Psib-lim-final}), holds now. Even if the beam 
from the phased array does not intercept the time reversal 
mirror at all, we will still get a time reversed signal 
at the source but with a much diminished amplitude. 
If the beam falls entirely within the time reversal mirror 
then the time reversed pulse will be a distorted form of 
$g(-t)$, with its amplitude reduced by the factor $(a/a_e)^d$. 
An interesting and important application of the time reversal 
of a beam in a random medium is the possibility of  
{\bf estimating} the effective aperture $a_e$ by pointing 
the beam in different directions toward the 
time reversal mirror, measuring the time 
reversed signal that back propagates to the source, that is, to the phased array, 
and inferring $a_e$ by fitting the measurements to (\ref{Psib-lim-final}).

\section{Summary and conclusions}

We have analyzed and explained two important phenomena associated with 
time reversal in a random medium:  
\begin{itemize} 
\item{Super-resolution of the 
back-propagated signal due to multipathing} 
\item{Self-averaging 
that gives a statistically stable 
refocusing} 
\end{itemize} 
Our analysis is based on a specific asymptotic limit 
(see Section \ref{sec:scaling}) where the 
longitudinal distance of propagation is much larger than the size of 
the time-reversal mirror, which in turn is much larger than the 
correlation length of the medium, fluctuations in the 
index of refraction are weak, and the wave length 
is short compared to the correlation length.  
This asymptotic regime is more relevant to optical or infrared 
time reversal than it is to sonar or ultrasound. 
We have related the self-averaging 
properties of the back-propagated signal to those of functionals of 
the Wigner distribution.  
Self-averaging of these functionals implies 
the statistical stability of the time-reversed and back-propagated signal 
in the frequency domain, provided that the source function is not 
too broad compared to the Fresnel number (\ref{fresnel}). 
Time reversal refocusing of waves emitted from a distributed source 
is self-averaging only in the time domain.  
 
We apply our theoretical results about stochastic 
Wigner distributions to time reversal and discuss in detail 
super-resolution and statistical stability in 
section \ref{sec:TR}.
 
\begin{appendix}

\section{The white noise limit and the parabolic approximation} 
\label{add2} 
 
We collect here some comments on the scaling analysis of 
section \ref{sec:scaling} and refer to \cite{jpf,NW91,White} 
for additional comments and results on scaling and asymptotics  
in the high-frequency and white-noise regime. 
 
The dimensionless parameters $\delta,\eps,\gamma$  introduced by 
(\ref{dimensionless-parameters}) in Section \ref{sec:scaling}, 
along with the Fresnel number $\theta$ defined by (\ref{fresnel}), 
lead to the scaled parabolic wave equation (\ref{eq:parabolic-rescaled}). 
If we do not make the parabolic approximation and keep the $\psi_{zz}$ 
term we have the scaled Helmholtz equation, with the phase $e^{ikz}$ removed, 
\begin{eqnarray} 
  \label{eq:full-rescaled} 
\frac{\eps^2\theta^2}{\delta^2} \psi_{zz}+ 2ik\theta\psi_z+\theta^2\Delta_\vx\psi+ 
\frac{k^2\delta}{\eps^{1/2}}\mu(\frac{\vx}{\delta},\frac{z}{\eps})\psi=0. 
\end{eqnarray} 
Here, as in (\ref{eq:parabolic-rescaled}), we relate the strength 
of the fluctuations $\sigma$ to $\eps$ and $\delta$ by (\ref{parameter-eps}). 
Is the parabolic approximation 
valid in the ordering (\ref{order-2}), $\theta \ll \eps \ll  \delta \ll 1$, 
 that we have analyzed? The answer is yes but not before both $\theta$ 
{\bf and} $\eps$ limits have been taken, in which case the scaled 
Wigner distribution (\ref{eq:wig-def}) converges to the Liouville-Ito 
process that is defined by the stochastic partial differential  
equation (\ref{eq:ito-liouville}).  
 
It is in the white noise limit $\eps\to 0$, with Fresnel number 
$\theta$ and $\delta$ fixed, that the parabolic approximation is valid for 
(\ref{eq:full-rescaled}), as was pointed out in \cite{jpf}. 
This is easily seen if the random fluctuations $\mu$ are 
differentiable in $z$.  
The parabolic approximation is clearly not valid in the high 
frequency limit $\theta\to 0$, before the white noise limit 
$\eps\to 0$ is also taken. 
In the white noise limit, the wave function $\psi(z,\vx)$ satisfies an 
Ito-Schr\"odinger equation  
\be 
\label{ito-schroedinger} 
2ik\theta d_z \psi +\theta^2\Delta_\vx\psi dz +  
\frac{ik^3 \delta^2}{4\theta}R_0(0)\psi dz +k^2\delta \psi d_zB(\frac{\vx}{\delta},z)=0 . 
\ee 
Here $R_0$ is the integrated covariance of the 
fluctuations $\mu$ given by (\ref{diff-coeff}) and (\ref{rzero}), and 
 the Brownian field 
$B(\vx,z)$ has covariance 
\[ 
\langle B(\vx,z_1)B(\vy,z_2) \rangle = R_0(\vx-\vy) z_1\wedge z_2 . 
\] 
This Ito-Schr\"odinger equation is the result of the central limit theorem
applied to (\ref{eq:full-rescaled}). Let
\[
B^{\eps}(\vx,z) =\frac{1}{\sqrt{\eps}}\int_0^z \mu(\vx,\frac{s}{\eps}) ds.
\]
Then, as $\eps\to 0$ this process converges weakly, under suitable hypotheses, to the
Brownian field $B(\vx,z)$ with the above covariance. The extra term in
(\ref{ito-schroedinger}) is the Stratonovich correction. 

The white noise limit for stochastic partial differential equations
is analyzed in \cite{BP84} and a rigrous theory of the Ito-Schr\"odinger
equation is given in \cite{Dawson-Papanico}. The ergodic theory of
the Ito-Schroedinger equation is explored in \cite{FPS98}. Wave propagation
in the parabolic approximation with white-noise fluctuations is considered
in detail in \cite{F93,TIV93}.

The scaled Wigner distribution for the process $\psi$, defined by 
(\ref{eq:wig-def}), satisfies the stochastic transport 
equation 
\begin{eqnarray} 
\label{W-st-transport} 
&& d_z W_{\theta}(z,\vx,\bp) + \frac{\bp}{k}\cdot \nabla_x W_{\theta}(z,\vx,\bp)dz \\ 
&&= \frac{k^2 \delta^2}{4\theta^2} \int \frac{d\bq}{(2\pi)^d} 
\hat{R}_0 (\bq) \left( W_{\theta}(z,\vx,\bp +\frac{\theta \bq}{\delta}) 
-W_{\theta}(z,\vx,\bp)\right)dz \nonumber \\ 
&&+ \frac{ik\delta}{2\theta}\int \frac{d\bq}{(2\pi)^d} e^{i\bq\cdot \vx/\delta} 
\left( W_{\theta}(z,\vx,\bp -\frac{\theta \bq}{2\delta}) - 
W_{\theta}(z,\vx,\bp +\frac{\theta \bq}{2\delta})\right)d_z \hat{B}(\bq,z) ,\nonumber 
\end{eqnarray} 
which is derived from (\ref{ito-schroedinger}) equation using the Ito calculus.
The Wigner process $W_{\theta}$ converges in the limit $\theta\to 0$ 
to the Liouville-Ito process defined by the stochastic partial 
differential equation (\ref{eq:ito-liouville}).

\newpage 
 
\section{Decorrelation of the Wigner process} 
\label{app-c} 
 
\subsection{Proof of Theorem \ref{lem1}} 
We give here the proof of 
Theorems \ref{lem1} and \ref{lem2}. 
We consider Theorem \ref{lem1} first. 
It will follow from the Lebesgue 
dominated convergence theorem if we show that for 
$\bp_1\ne \bp_2$: 
\begin{equation}\label{decorrel} 
E\left\{W_\delta(z,\vx,\bp_1)W_\delta(z,\vx,\bp_2)\right\} 
-E\left\{W_\delta(z,\vx,\bp_1)\right\}E\left\{W_\delta(z,\vx,\bp_2)\right\} 
\to 0 
\end{equation} 
as $\delta\to 0$ because the function $W_\delta$ is uniformly bounded 
and $E\left\{W_\delta(z,\vx,\bp_1)\right\}$ does not depend on 
$\delta$.  Furthermore, the correlation function at the same spatial 
point but for two different values of the wave vector, 
$U_\delta^{(2)}(z,\vx,\bp_1,\bp_2)= 
E\left\{W_\delta(z,\vx,\bp_1)W_\delta(z,\vx,\bp_2)\right\}$ is the 
solution of (\ref{eq:difflimit-N}) with $N=2$ and the initial data 
\[ 
W_\delta^{(2)}(0,\vx_1,\bp_1,\vx_2,\bp_2)=W_I(\vx_1,\bp_1)W_I(\vx_2,\bp_2), 
\] 
evaluated at $\vx_1=\vx_2=\vx$. 
Therefore $U_\delta^{(2)}$ may be represented as 
\[ 
U_\delta^{(2)}(z,\vx_1,\bp_1,\vx_2,\bp_2)= 
E\left\{W_I(\vX_\delta^1(z),\bP_\delta^1(z)) 
W_I(\vX_\delta^2(z),\bP_\delta^2(z))\right\}. 
\] 
The processes $\vX_\delta^{1,2}$ and $\bP_\delta^{1,2}$ satisfy the system of 
SDE's (\ref{Npoint-sde}) which may be more explicitly written as 
\begin{eqnarray}\label{delta-sys} 
&&d\bP_\delta^1=-\left[\sigma(0) d{\bf B}^1(z)+ 
\frac 12 \sigma\left(\frac{\vX_\delta^1-\vX_\delta^2}{\delta}\right)d\vB^2(z) 
\right]\\ 
&&d\bP_\delta^2=-\left[\sigma(0) d\vB^2(z)+ 
\frac 12 \sigma\left(\frac{\vX_\delta^2-\vX_\delta^1}{\delta}\right)d\vB^1(z) 
\right]\nonumber\\ 
&&d\vX_\delta^1=-\bP_\delta^1dz, ~~ 
d\vX_\delta^2=-\bP_\delta^2dz\nonumber 
\end{eqnarray} 
with the initial conditions 
$\vX_\delta^{1,2}(0)=\vx,~~\bP_\delta^m(0)=\bp_m$, $m=1,2$.  Here 
$\sigma^2(0)=D$, the diffusion coefficient (\ref{diff-coeff}) and the 
coupling matrix $\sigma(\vx)$ is given by (\ref{multi-cov}). 
Recall that $W_\delta(z,\vx,\bp,k)=W_\delta(z,\vx,\bp/k;1)$ 
and we need only consider the case $k=1$. 
 
It is convenient to introduce the processes $\vX^{1,2}$ and 
$\bP^{1,2}$ that are solutions of (\ref{Npoint-sde}) with no coupling: 
\begin{eqnarray}\label{sys-uncoupl} 
&&d\bP^m=-\sigma(0) d{\bf B}^m(z),~~ 
d\vX^m=-\bP^mdz,\\ 
&&\vX^{1,2}(0)=\vx,~~\bP^m(0)=\bp_m, ~~m=1,2\nonumber 
\end{eqnarray} 
and define the deviations of the solutions of the coupled system of 
SDE's (\ref{delta-sys}) from those of (\ref{sys-uncoupl}):  
$\vZ_\delta^m=\vX_\delta^m-\vX^m$, $\bS_\delta^m=\bP_\delta^m-\bP^m$. 
Then we have 
\begin{eqnarray}\label{delta-sys1} 
&&d\bS_\delta^1=-\frac 12 
\sigma\left(\frac{\vX_\delta^1-\vX_\delta^2}{\delta}\right)d\vB^2(z),~~ 
d\bS_\delta^2=-\frac 12 
 \sigma\left(\frac{\vX_\delta^2-\vX_\delta^1}{\delta}\right)d\vB^1(z)\\ 
&&d\vZ_\delta^1=-\bS_\delta^1dz, ~~ 
d\vZ_\delta^2=-\bS_\delta^2dz\nonumber 
\end{eqnarray} 
with the initial data $\bS_\delta^m(0)=\vZ^m(0)=0$. 
Define
\begin{eqnarray}\label{decorrel0} 
&&
{\cal V}(\vX^1,\vX^2,\bP^1,\bP^2,
\vZ^1_\delta,\vZ^2_\delta,\bS^1_\delta,\bS^2_\delta)
\\ && \nonumber  =
W_I(\vX^1+\vZ_\delta^1,\bP^1+\bS_\delta^1) 
W_I(\vX^2+\vZ_\delta^2,\bP^2+\bS_\delta^2) 
-W_I(\vX^1,\bP^1)W_I(\vX^2,\bP^2)
\end{eqnarray} 
then we have  
with the above notation 
\begin{eqnarray}\label{decorrel1} 
&&E\left\{W_\delta(z,\vx,\bp_1)W_\delta(z,\vx,\bp_2)\right\} 
-E\left\{W_\delta(z,\vx,\bp_1)\right\}E\left\{W_\delta(z,\vx,\bp_2)\right\}\\ 
&&= 
E\left\{
{\cal V}(\vX^1(z),\vX^2(z),\bP^1(z),\bP^2(z),
\vZ^1_\delta(z),\vZ^2_\delta(z),\bS^1_\delta(z),\bS^2_\delta(z))
\right\}\nonumber\\ 
&& 
\le CE\left\{|\vZ_\delta^1(z)|+|\vZ_\delta^2(z)|+|\bS_\delta^1(z)|+ 
|\bS_\delta^2(z)| 
\right\}\nonumber 
\end{eqnarray} 
since $W_I$ is a Lipschitz function.  
 
Let us assume for simplicity 
that the correlation function $R(\vx)$ has compact support inside the 
set $|\vx|\le M$. Then the coupling term in (\ref{delta-sys}) is 
non-zero only when $|\vX_\delta^1-\vX_\delta^2|\le M\delta$.  We 
introduce the processes $\bQ_\delta=\bP_\delta^1-\bP_\delta^2$ and 
$\vY_\delta=\vX_\delta^1-\vX_\delta^2$ that govern 
(\ref{delta-sys1}). They satisfy the SDE's 
\begin{eqnarray}\label{qy-sys} 
&&d\bQ_\delta=-\left[\sigma(0) - 
\frac 12 \sigma\left(\frac{\vY_\delta}{\delta}\right)\right] 
d\tilde\vB,~~d\vY_\delta=-\bQ_\delta dz,\\ 
&&~~ 
\bQ_\delta(0)=\bp_1-\bp_2,~\vY_\delta(0)=0\nonumber 
\end{eqnarray} 
with $\tilde\vB=\vB^1-\vB^2$ being a Brownian motion.

In order to prove the theorem we show that the coupling term 
$\sigma(\cdot)$ in (\ref{delta-sys}) introduces only lower order 
correction terms, that is, $\bS_{\delta}^m$ and $\bZ_{\delta}^m$ are 
small.  We show first that after a small `time', $\tau$, the points 
$\vX_\delta^m$ are driven apart since 
$\bQ_\delta(0)=\bP_\delta^1(0)-\bP_\delta^2(0)\neq0$.  Then we show 
that after the points have separated the probability that they come 
close, so that the coupling term $\sigma(\cdot)$ becomes non-zero, is 
small.  This ``non-recurrence'' condition requires that the spatial 
dimension $d\ge 2$. It follows that to leading order the points 
$\vX_\delta^m$ are uncorrelated when $d\ge 2$ and that the coupling 
term introduces only lower order corrections. A similar argument for 
$d=1$ would require an estimate on the time that points that are 
originally separated in the spatial variable spend near each other, where 
the coupling term in (\ref{delta-sys}) is not zero. 
 
We need the 
following two Lemmas.  The first one shows that particles that start at 
the same point $\vx$ with different initial directions $\bp_1$ and 
$\bp_2$, get separated with a large probability: 
\begin{lemma}\label{lemma3} 
Let $\vY_\delta$, $\bQ_\delta$ solve (\ref{qy-sys}) with 
$\vY_\delta(0)=0$, $\bQ_\delta(0)=\bq\ne 0$. Then for any 
$\eps>0$ there exists $\tau_0(\eps)>0$ that depends only on 
$\bq=\bp_1-\bp_2$ but not on $\delta$ so that we have 
$\displaystyle P\left(|\vY_\delta(\tau)|\ge \frac{|\bq|\tau}{2}\right)\ge  
1-\eps$ for all $\tau\le\tau_0(\eps)$. 
\end{lemma} 
 
The second lemma shows that after the particles are separated, the 
probability that they come close to each other is small: 
\begin{lemma}\label{lemma4}  
Given any fixed $r>0$ and $z>0$, if $\vY_\delta$, $\bQ_\delta$ solve 
(\ref{qy-sys}) with $|\vY_\delta(0)|\ge r$, $\bQ_\delta(0)=\bq\ne 0$, 
then $P\left(\inf_{0\le s\le z} |\vY_\delta(s)|\le M\delta\right)\to 
0$ as $\delta\to 0$. 
\end{lemma} 
 
We prove Theorem \ref{lem1} before proving Lemmas \ref{lemma3} and 
\ref{lemma4}: 
\begin{proof}
Let $z$ and $\bq=\bp_1-\bp_2$ be fixed and defined as above.
Given $\eps>0$, then for any $\tau<\tau_0(\eps)$ (with $\tau_0$ as defined
in Lemma \ref{lemma3}), 
Lemma \ref{lemma4} and the Markov property of 
the Brownian motion imply that 
\[ 
P\left(\bS_\delta^m(z)=\bS_\delta^m(\tau) 
\Big||\vY_\delta(\tau)|\ge \frac{\tau|\bq|}{2} \right) 
\ge 1-\eps  
\] 
and 
\[ 
P\left(\vZ_\delta^m(z)=\vZ_\delta^m(\tau)+(z-\tau)\bS_\delta^m(\tau) 
\Big||\vY_\delta(\tau)|\ge \frac{\tau|\bq|}{2}\right) 
\ge 1-\eps 
\] 
for $\delta<\delta_0(\tau,\eps)$. Furthermore, 
\begin{eqnarray}\label{taubd} 
& &
E\left\{|\vZ_\delta^1(\tau)|+|\vZ_\delta^2(\tau)|+|\bS_\delta^1(\tau)|+ 
|\bS_\delta^2(\tau)|\Big||\vY_\delta(\tau)|\ge \frac{\tau|\bq|}{2} 
\right\} \\
&\leq&
E\left\{|\vZ_\delta^1(\tau)|+|\vZ_\delta^2(\tau)|+|\bS_\delta^1(\tau)|+ 
|\bS_\delta^2(\tau)| \right\}/(1-\eps)
\le C\tau  \nonumber
\end{eqnarray} 
because the function $\sigma$ is uniformly bounded. Therefore we have 
\begin{eqnarray} 
&&~~~~~~~~~E\left\{
{\cal V}(\vX^1,\vX^2,\bP^1,\bP^2,
\vZ^1_\delta,\vZ^2_\delta,\bS^1_\delta,\bS^2_\delta)
\right\}\nonumber\\ 
&&~~~~~~~~~= 
E\left\{
{\cal V}(\vX^1,\vX^2,\bP^1,\bP^2,
\vZ^1_\delta,\vZ^2_\delta,\bS^1_\delta,\bS^2_\delta)
\Big||\vY_\delta(\tau)|\ge \frac{\tau|\bq|}{2}\right\} 
P\left(|\vY_\delta(\tau)|\ge \frac{\tau|\bq|}{2}\right)
\label{1+2}\\ 
&&+ 
E\!\left\{
{\cal V}(\vX^1,\vX^2,\bP^1,\bP^2,
\vZ^1_\delta,\vZ^2_\delta,\bS^1_\delta,\bS^2_\delta)
\Big||\vY_\delta(\tau)|\le\frac{\tau|\bq|}{2}\right\} 
P\left(\!|\vY_\delta(\tau)|\le \frac{\tau|\bq|}{2}\right)=I+II.\nonumber 
\end{eqnarray} 
The second term above is small because the probability for 
$\vY_\delta(\tau)$ to be very small is bounded by Lemma \ref{lemma3}. 
More precisely, given $\eps>0$ and $\tau < \tau_0(\eps)$,
Lemma \ref{lemma3} implies that
\begin{equation} 
\label{IIbd} 
II\le C\eps. 
\end{equation} 
 
The first term in (\ref{1+2}) corresponds to the more 
likely scenario that $\vY_\delta$ at time $\tau$ has left the ball of radius 
$\tau|\bq|/2$.
We estimate it as 
follows.  The probability that $\vY_\delta$ re-enters the ball of radius 
$M\delta$ is small according to Lemma \ref{lemma4}. Moreover, if $\vY_\delta$ 
stays outside this ball, the difference variables $\vZ^m$ and $\bS^m$ 
are bounded in terms of their values at time $\tau$.  The latter are 
small if $\tau$ is small.  More precisely, using (\ref{taubd}) we  
choose $\tau$ so small that 
\[ 
E\left\{|\vZ_\delta^1(\tau)|+|\vZ_\delta^2(\tau)|+|\bS_\delta^1(\tau)|+ 
|\bS_\delta^2(\tau)|\Big||\vY_\delta(\tau)|\ge \frac{\tau|\bq|}{2}\right\} 
\le \eps. 
\] 
Then we obtain 
\begin{eqnarray*} 
&&I\le E\left\{
{\cal V}(\vX^1,\vX^2,\bP^1,\bP^2,
\vZ^1_\delta,\vZ^2_\delta,\bS^1_\delta,\bS^2_\delta)
\Big||\vY_\delta(\tau)|\ge \frac{\tau|\bq|}{2}\right\} \\ 
&&\le E\left\{
{\cal V}(\vX^1,\vX^2,\bP^1,\bP^2,
\vZ^1_\delta,\vZ^2_\delta,\bS^1_\delta,\bS^2_\delta)
\Big||\vY_\delta(\tau)|\ge \frac{\tau|\bq|}{2}\hbox{ and } 
\inf_{\tau\le s\le z} |\vY_\delta(s)|\le M\delta\right\}\\ 
&&\times P\left(\inf_{\tau\le s\le z} |\vY_\delta(s)|\le  
M\delta\Big||\vY_\delta(\tau)|\ge \frac{\tau|\bq|}{2}\right)\\ 
&&+ 
E\left\{|\vZ_\delta^1(z)|+|\vZ_\delta^2(z)|+|\bS_\delta^1(z)|+ 
|\bS_\delta^2(z)|\Big||\vY_\delta(\tau)|\ge \frac{\tau|\bq|}{2}\hbox{ and } 
\inf_{\tau\le s\le z} |\vY_\delta(s)|\ge M\delta\right\}\\ 
&&\times P\left(\inf_{\tau\le s\le z} |\vY_\delta(s)|\ge  
M\delta\Big||\vY_\delta(\tau)|\ge \frac{\tau|\bq|}{2}\right)=I_1+I_2.  
\end{eqnarray*} 
The term $I_1$ goes to zero as $\delta\to 0$ by Lemma \ref{lemma4}.
However, if the conditions in $I_2$ hold, then 
\[ 
\bS_\delta^m(z)=\bS_\delta^m(\tau), ~~ 
\vZ_\delta^m(z)=\vZ_\delta^m(\tau)-\frac 1k(z-\tau)\bS_\delta^m(\tau). 
\] 
Therefore the term $I_2$ may be bounded with the help of (\ref{taubd}) 
by 
\begin{eqnarray*} 
&&I_2\le E\left\{|\vZ_\delta^1(z)|+|\vZ_\delta^2(z)|+|\bS_\delta^1(z)|+ 
|\bS_\delta^2(z)|\Big||\vY_\delta(\tau)|\ge \frac{\tau|\bq|}{2}\hbox{ and } 
\inf_{\tau\le s\le z} |\vY_\delta(s)|\ge M\delta\right\}\\ 
&&\le CE\left\{|\vZ_\delta^1(\tau)|+|\vZ_\delta^2(\tau)|+|\bS_\delta^1(\tau)|+ 
|\bS_\delta^2(\tau)|\Big||\vY_\delta(\tau)|\ge \frac{\tau|\bq|}{2}\right\} 
\le C\tau. 
\end{eqnarray*} 
Putting together (\ref{1+2}), (\ref{IIbd}) and the above bounds on 
$I_1$ and $I_2$, we obtain 
\[ 
E\left\{|\vZ_\delta^1(z)|+|\vZ_\delta^2(z)|+|\bS_\delta^1(z)|+ 
|\bS_\delta^2(z)|\right\}\le C\eps 
\] 
for $\delta<\bar{\delta}$ and Theorem \ref{lem1} follows from (\ref{decorrel1}). 
\end{proof} 
 
\subsection{Proof of Lemmas \ref{lemma3} and \ref{lemma4}} 
We first prove Lemma \ref{lemma3}. 
\begin{proof} 
We write 
\[ 
Q_\delta(z)=\bq -\int_0^z \left(\sigma(0)-\frac{1}{2} 
\sigma(\vY(s)/\delta)\right) ~d\tilde\vB(s) ~\equiv~ \bq +\tilde\bQ_\delta(z) 
\] 
so that 
\[ 
\vY_\delta(t)=-\bq t-\int_0^t\tilde\bQ_\delta(s)ds.  
\] 
Then we have 
\begin{eqnarray} 
\label{res1.1} 
P\left( \sup_{0 \leq s \leq \tau}  
|\tilde\bQ_\delta(s)|~>~ r \right)\leq C \tau/r^2 
\end{eqnarray} 
and hence 
\[ 
P\left(|\vY_\delta(\tau)+\tau\bq|>r\tau\right)\le 
P\left( \sup_{0 \leq s \leq \tau}|\tilde\bQ_\delta(s)|>r \right)  
\le C \tau/r^2.  
\] 
We let $r=|\bq|/2$ in the above formula and obtain 
\[ 
P\left(|\vY_\delta(\tau)|<\frac{\tau|\bq|}{2}\right)\le\frac{C}{|\bq|^2}\tau, 
\] 
and the conclusion of Lemma \ref{lemma3} follows. 
\end{proof} 
 
Finally, we prove Lemma \ref{lemma4}.  
\begin{proof} 
Let $\tau_\delta$ be the first time $\vY_\delta(z)$ enters the ball of radius 
$M\delta$: 
\[ 
\tau_\delta=\inf\left\{z:~|\vY_\delta(z)|\le M\delta\right\}, 
\]
with $\vY_\delta(0)=\vY^0 \neq 0$. 
For $0<\alpha<1$ let $\Delta z=\delta^{1-\alpha}$, $n=\lceil z/\Delta 
z \rceil$, $J_i=(i\Delta z,(i+1)\Delta z)$ and $p<1$.  Note that until 
the time $\tau_\delta$ the process $(\vY_\delta,\bQ_\delta)$ coincides 
with the process $(\vY,\bQ)$ governed by (\ref{qy-sys}) without the
coupling term
$\sigma(\vY_\delta/\delta)$.
We  find 
\begin{eqnarray*} 
 P(\tau_\delta<z) \le \sum_{i=0}^{n-1} \left\{ 
 P\left(|\vY(i\Delta z)| < M \delta^p\right) + 
 P\left(\inf_{s\in J_i}\! |\vY(s)| < M \delta 
  ~\Big|~ |\vY(i\Delta z)| \geq M\delta^p\right) \right\} . 
\end{eqnarray*} 
The process $\vY(s)$ is  Gaussian with mean $\vY^0$ 
and variance ${\cal O}(s^2)$. Therefore, 
there is a  $ \bar{\delta} >0$
such that
for $\delta < \bar{\delta}$  
\begin{eqnarray*} 
 P(|\vY(i\Delta z)| < M \delta^p) &\leq&  C \delta^{d p}   . 
\end{eqnarray*} 
If we assume
\begin{eqnarray}
\label{pcond}
p < 1-\alpha
\end{eqnarray} 
then also
\begin{eqnarray*} 
 P(\tau_\delta<z) &\le& 
 n C \left(  \delta^{d p} + 
 P\left(\sup_{0<s<\Delta z} |\vY(s)-\vY^0| \geq 
 M [\delta^p-\delta] \right)
\right)  \\ 
&\leq&
C ( \delta^{dp+\alpha-1} + 
     \delta^{\alpha-1}
 P\left(\sup_{0<s<\Delta z} |\vB(s)| \geq 
 M [\delta^p-\delta]/ \Delta z\right)
) \\
&\leq&
C ( \delta^{dp+\alpha-1} + 
     \delta^{\alpha-1}
\frac{E\left\{ \vB(\Delta z)^{2r}\right\} \Delta z^{2r} } 
              {(\delta^p-\delta)^{2r} }
) \\
&\leq&
 C\left[ \delta^{dp+\alpha-1}+\delta^{\alpha-1-rp+3r(1-\alpha)/2}\right].
\end{eqnarray*} 
Note that with $p<1-\alpha$ and $r$  large enough, there is
a $q>0$ so that 
\[ 
 P(\tau_\delta<z) \le C\delta^{q}  
\]
if $d \geq 2$
and Lemma \ref{lemma4} follows. 
\end{proof}

\subsection{Proof of Theorem \ref{lem2}} 
 
We need to show first that  
\be 
\label{fin-int} 
J_{\delta}(z,\vx)= \int W_{\delta}(z,\vx,\bp) d\bp 
\ee 
is finite with probability one. The stochastic flow 
$(\vX_{\delta}(t,\vx,\bp),\bP_{\delta}(t,\vx,\bp)$ 
is continuous in $(t,\vx,\bp)$ with probability one, 
so $W_{\delta}(z,\vx,\bp)=W_I(\vX_{\delta}(t,\vx,\bp),\bP_{\delta}(t,\vx,\bp))$ 
is bounded and continuous. It is, moreover, non-negative if 
$W_I \geq 0$. We know that 
\[ 
\int E\{ W_{\delta}(z,\vx,\bp) \}d\bp 
\] 
is finite and independent of $\delta$, and the order of integration and 
expectation can be interchanged by Tonelli's theorem. This theorem 
implies in addition that $J_{\delta}(z,\vx)$ is finite with probability 
one. 
 
We can now consider  
\[ 
E\{J_{\delta}^2 (z,\vx)\} =  
\int E\{ W_{\delta}(z,\vx,\bp_1) W_{\delta}(z,\vx,\bp_2)\} 
d\bp_1 d\bp_2  . 
\] 
The integrand is bounded by an integrable function uniformly in 
$\delta$ because 
\[ 
E\{ W_{\delta}(z,\vx,\bp_1) W_{\delta}(z,\vx,\bp_2)\} 
\leq E^{1/2}\{ W_{\delta}^2(z,\vx,\bp_1)\} E^{1/2}\{ W_{\delta}(z,\vx,\bp_2)\} , 
\] 
the right side does not depend on $\delta$, and is integrable. 
Therefore by the Lebesgue dominated convergence theorem and the  
results of the previous Section we have that 
\[ 
\lim_{\delta\to 0}E\{J_{\delta}^2 (z,\vx)\} = 
E^2 \{J_{\delta}(z,\vx)\} 
\] 
and the right side does not depend on $\delta$. This completes the 
proof of Theorem \ref{lem2}.

\end{appendix}


\begin{thebibliography}{99} 
 
\bibitem{jpf} 
F. Bailly, J.F. Clouet and J.P. Fouque, 
Parabolic and white noise approximation for waves in random media, 
SIAM Journal on Applied Mathematics {\bf 56}, 1996, 1445-1470. 
 
\bibitem{Bal-Ryzhik1} 
G.Bal and L. Ryzhik, Time reversal for classical waves in random media, 
Comptes rendus de l'Acad\'emie des sciences - S\'erie I -  
Math\'ematique, {\bf 333}, 2001, 1041-1046. 
 
\bibitem{Bal-Ryzhik2} 
G.Bal and L. Ryzhik, Time reversal for waves in random media,  
Preprint, 2002. 
 
\bibitem{BPR} 
G.Bal, G. Papanicolaou and L. Ryzhik, Self-averaging in time reversal 
for the parabolic wave equation Preprint, 2002. 
 
\bibitem{BPR-Nonlin} 
G.Bal, G. Papanicolaou and L. Ryzhik, Radiative transport limit 
for the random Schr\"odinger equation, Nonlinearity, {\bf 15}, 2002, 513-529. 
 
 
\bibitem{BP78} 
G. Blankenship and G. C. Papanicolaou, 
Stability and Control of Stochastic Systems with Wide-Band Noise Disturbances, 
SIAM J. Appl. Math., 
{\bf 34}, 
1978, 
437-476. 
 
 
\bibitem{BPZ} P.~Blomgren, G.~Papanicolaou, and H.~Zhao,  
Super-Resolution in Time-Reversal Acoustics, J. Acoust. Soc. Am., {\bf 
111}, 2002, 230-248. 
   
\bibitem{BTPB-1} L. Borcea, C. Tsogka, G. Papanicolaou and J. Berryman, 
Imaging and time reversal in random media, to appear in Inverse Problems, 2002. 
 
\bibitem{BTPB-2}  J. Berryman, L. Borcea, G. Papanicolaou and C. Tsogka,  
Statistically stable ultrasonic imaging in random media, 
Preprint, 2002. 

\bibitem{BP84}
R. Bouc and E. Pardoux,
Asymptotic analysis of PDEs with wide-band noise
 disturbances and expansion of the moments,
Stochastic Analysis and Applications,
{\bf 2}, 1984, 369-422.

 
\bibitem{Dawson-Papanico} D. Dawson and G. Papanicolaou,  
A random wave process, Appl. Math. Optim., {\bf 12}, 1984, 97--114.   
 
\bibitem{DJ90} D. Dowling and D. Jackson, Phase conjugation in underwater 
    acoustics, Jour. Acoust. Soc. Am., {\bf 89}, 1990, 171-181 
     
\bibitem{DJ92} D. Dowling and D. Jackson, Narrow-band performance of 
    phase-conjugate arrays in dynamic random media, Jour. Acoust. Soc. 
    Am.,{\bf 91}, 1992, 3257-3277. 
 
 
\bibitem{Fink-Nonlin} 
M. Fink and J. de Rosny, Time-reversed acoustics in random media and 
in chaotic cavities, Nonlinearity, {\bf 15}, 2002, R1-R18. 
 
\bibitem{FCDPRTTW} 
M. Fink, D. Cassereau, A. Derode, C. Prada, P. Roux, M. Tanter, 
J.L. Thomas and F. Wu, Time-reversed acoustics, Rep. Progr. Phys., 
{\bf 63}, 2000, 1933-1995. 
 
\bibitem{Fink-Prada-01} 
{M.~Fink and C.~Prada}, {{Acoustic time-reversal mirrors}}, Inverse 
  Problems, {\bf 17}, 2001, R1--R38. 
  
   
 
\bibitem{FPS98} 
J.P. Fouque, G.C. Papanicolaou and Y. Samuelides,  
Forward and Markov Approximation: The Strong Intensity Fluctuations Regime Revisited,  
Waves in Random 
Media, {\bf 8}, 1998, 303-314.  
 
\bibitem{F93}
K. Furutsu,
Random Media and Boundaries: Unified Theory, Two-Scale
 Method, and Applications,
Springer Verlag, 1993.

  
\bibitem{GMMP} 
P.G\'erard, P.Markovich, N.Mauser and F.Poupaud, Homogenization limits 
 and Wigner transforms, Comm.Pure Appl. Math., {\bf 50}, 1997, 
 323-380. 
 
\bibitem{HSK99} W. Hodgkiss, H. Song, W. Kuperman, T. Akal, C. Ferla 
  and D. Jackson, A long-range and variable focus phase-conjugation 
  experiment in a shallow water, Jour. Acoust. Soc. Am., {\bf 105}, 
  1999, 1597-1604. 
 
 
\bibitem{KP79}  
H. Kesten and G. Papanicolaou,  
A Limit Theorem for Turbulent Diffusion, 
Comm. Math. Phys., {\bf 65}, 1979, 97-128. 
 
 
\bibitem{Kohler-Papanico} G. Papanicolaou and W. Kohler, Asymptotic 
analysis of deterministic and stochastic equations with rapidly 
varying components, Comm. Math. Phys. {\bf 45}, 217--232, 1975. 
 
\bibitem{Kunita} H. Kunita, {\it Stochastic flows and stochastic 
differential equations}. Cambridge 
Studies in Advanced Mathematics, 24. Cambridge University Press, 
Cambridge, 1997. 
 
\bibitem{KHS97} W. Kuperman, W. Hodgkiss, H. Song, T. Akal, C. Ferla 
  and D. Jackson, Phase-conjugation in the ocean, Jour. Acoust. Soc. 
  Am., {\bf 102}, 1997, 1-16. 
 
\bibitem{KHSAFJ} 
W. Kuperman, W. Hodgkiss, H. Song, T. Akal, C. Ferla and D. Jackson, 
Phase conjugation in the ocean: Experimental demonstration of an acoustic 
time reversal mirror, J. Acoust. Soc. Am., {\bf 103}, 1998, 25-40. 
 
\bibitem{K84} 
H. Kushner, 
Approximation and weak convergence methods for random processes,  
with applications to stochastic systems theory, 
MIT Press Series in Signal Processing, Optimization, and Control,  
MIT Press, 
1984. 
 
 
\bibitem{NW91} 
B. Nair and B. White, High-frequency  wave propagation 
in random media- a unified approach, 
SIAM J. Appl. Math., {\bf 51}, 1991, 374-411. 
 
 
\bibitem{RPK-WM}  L.~Ryzhik, G.~Papanicolaou, and J.~B. Keller. 
{Transport equations for elastic and other waves in random media}, 
Wave Motion, {\bf 24}, 327--370, 1996. 
 
 
\bibitem{Tappert} F. Tappert, The parabolic approximation method, 
  Lecture notes in physics, vol. 70, {\it Wave propagation and underwater 
  acoustics}, Springer-Verlag, 1977. 

\bibitem{TIV93}
V. I. Tatarskii, A. Ishimaru and V. U. Zavorotny, editors,
Wave Propagation in Random Media (Scintillation),
SPIE and IOP, 1993.

 
\bibitem{fink2} Thomas J. L. and M. Fink,  
{\em Ultrasonic beam focusing through tissue inhomogeneities 
with a time reversal mirror: Application to transskull therapy}, 
IEEE Trans. on Ultrasonics, Ferroelectrics and Frequency Control, 
Vol. 43, 1122-1129, (1996).  
 
\bibitem{TsP} C. Tsogka and G. Papanicolaou, 
Time reversal through a solid-liquid interface and super-resolution, 
Preprint, 2001. 
 
\bibitem{White} B. White, The stochastic caustic, SIAM Jour. Appl. 
  Math., {\bf 44}, 1984, 127-149. 
 
\end{thebibliography}
\end{document}